\documentclass[letterpaper]{article}
\usepackage{amsmath}
\usepackage{amssymb}
\usepackage[numbers,sort&compress]{natbib} 
\usepackage{alifeconf}
\usepackage{url,hyperref,cleveref}
\usepackage[superscript]{cite}
\usepackage{booktabs}
\usepackage{adjustbox}
\usepackage{xcolor}
\usepackage{enumitem}
\usepackage{epigraph}

\usepackage{xr}
\usepackage{tikz-cd}
\usepackage[normalem]{ulem}

\usepackage[draft,inline,nomargin]{fixme} \fxsetup{theme=color} 
\FXRegisterAuthor{cg}{cgg}{\color{red}CGG}

\title{ALIFE2024 template}

\title{Characterizing Open-Ended Evolution Through Undecidability Mechanisms in Random Boolean Networks}

\author{
    Amahury J. López-Díaz$^{1,*}$,
    Pedro Juan Rivera Torres$^{2,4}$,
    Gerardo L. Febres$^{1,3}$, \and
    Carlos Gershenson$^{1}$ \\
    \mbox{}\\
    $^1$School of Systems Science and Industrial Engineering, Binghamton University, 4400 Vestal Pkwy E, Binghamton, NY 13902, USA \\
    $^2$Department of Computer Science and Automation, University of Salamanca, 1 Patio de Escuelas, Salamanca, 37008, Spain \\
    $^3$Department of Processes and Systems, Universidad Simón Bolívar, Sartenejas, Baruta 1080, Miranda, Venezuela \\
    $^4$St. Edmund's College, University of Cambridge, Cambridge, CB3 0BN, United Kingdom\\
    $^*$Corresponding Author:
    alpez@binghamton.edu
} 

\begin{document}

\maketitle

\begin{abstract}
Discrete dynamical models underpin systems biology, but we still lack substrate-agnostic diagnostics for identifying finite-horizon dynamical signatures that may be relevant to open-ended evolution (OEE), such as the recurrent production of novel phenotypic states rather than rapid settling or unstructured noise. We introduce a simple, model-independent metric, $\Omega$, that summarizes the residence-time-weighted contribution of attractor cycle lengths across the sequence of recurrent episodes realized within a finite observation window. $\Omega$ is zero for single-attractor dynamics and also vanishes for pure novelty without recurrence, while increasing when trajectories repeatedly enter multiple persistent cyclic phenotypes. Using Random Boolean Networks (RBNs) as a controlled testbed, we compare classical Boolean dynamics with biologically motivated non-classical mechanisms (probabilistic context switching, annealed rule mutation, paraconsistent logic, modal necessary/possible gating, and quantum-inspired superposition/paired-state coupling) under homogeneous and heterogeneous updating schemes. Our results support the view that undecidability-adjacent, state-dependent mechanisms---implemented as probabilistic context switching, modal necessity/possibility gating, paraconsistent logic, or quantum-inspired correlated branching---are enabling conditions for sustained novelty. At the end of our manuscript we outline a practical extension of $\Omega$ to continuous/hybrid state spaces, positioning $\Omega$ as a portable proxy for OEE in biological modeling and a guide for engineering evolvable synthetic circuits.
\end{abstract}

\section{Introduction} 
Biological evolution has generated a broad diversity of forms and functions across scales, from microbes to large multicellular organisms.
As an example, if we focus on microbial species alone, their variety has been estimated to be $\sim$$10^{12}$,~\cite{locey2016scaling} which is approximately the number of galaxies in the observable universe.~\cite{conselice2016evolution} One of the central lessons from Darwin is that this diversity is not a static catalog, but is dynamically produced by evolution.~\cite{Darwin1859} Even though any real biosphere is finite and the underlying physical and chemical mechanisms have remained largely unchanged for roughly 4 billion years, the biosphere has continued to explore new combinations of forms and functions without any clear sign of saturation. In this practical sense, biological evolution appears \emph{effectively open-ended} on geological timescales~\cite{packard2019open}: the combinatorial space of possible genotypes and phenotypes is so large, and the dynamics of evolution so far from equilibrium, that ongoing novelty can be expected rather than exceptional. In the artificial life literature, this has motivated the study of \emph{open-ended evolution} as the ongoing production of novel, adaptive structures without an intrinsic upper bound on diversity or complexity.~\cite{packard2019overview} Since processes having the property of ongoing creative productivity are not necessarily biological, we have generalized this conceptualization to include \emph{open-endedness} in any form.~\cite{stanley2019open}

Despite its ubiquity in the literature, there is no full consensus on how to define open-ended evolution (OEE). Some authors take the continual generation of novelty---the sustained appearance of previously unseen entities or behaviors---to be sufficient for a system to be labeled as \emph{open-ended}.~\cite{soros2014identifying} Others argue that OEE should be reserved for cases in which some notion of system \emph{complexity} can grow without an a priori upper bound, as seems to be suggested by empirical scaling laws across levels of biological organization.~\cite{corominas2018zipf} However, this immediately raises the methodological problem of how to define and measure complexity in a way that is both domain-independent and monotonically increasing along evolutionary trajectories. As a result, the literature distinguishes between variety-based and complexity-based conceptions, and more generally between multiple ``flavors'' of open-endedness.~\cite{taylor2016open} In this manuscript we do not claim to resolve this definitional debate. Instead, we adopt a narrower operational target: the capability of a system to continually generate novel (without necessarily being increasingly complex) structures or behaviors without settling into fixed or cyclic states. This target captures a signature that is plausibly relevant to OEE, but should not be confused with a complete biological definition of OEE.

So far, two overarching suites of metrics have been proposed to measure OEE. On the one hand, \emph{Evolutionary Activity Statistics} (EAS) measure the degree of adaptive dynamics within a population by tracking the persistence of specific elements, called “components.”~\cite{bedau1997comparison, bedau1998classification} As a disadvantage of this approach, we should know what the “components” are \emph{ab initio}; these, in fact, are selected arbitrarily. Furthermore, it also fails to take into account that “components” can evolve over time, disregarding their ability to acquire new functions (or become obsolete) through adaptation.~\cite{wong2023roles} On the other hand, \emph{Measurements of Open-ended Dynamics in Evolving Systems} (MODES) is a toolbox designed to analyze and quantify the dynamics of evolving systems, specifically focusing on identifying features that are indicative of OEE.~\cite{dolson2019modes} Although MODES is an ambitious proposal, its application is limited to specific systems. In fact, some of its metrics only make sense in AVIDA, an artificial life software platform to study the evolutionary biology of self-replicating and evolving digital organisms.~\cite{lenski2003evolutionary}

Moreover, classifying a system as open-ended does not help us to understand how this property arose evolutionary speaking. As has been pointed out by Stepney and Hickinbotham,~\cite{stepney2024open} the focus should be on scrutinizing the mechanisms underlying open-endedness, not simply trying to quantify it. Of course this requires a thinking shift with respect the current approaches to OEE, suggesting that the capacity for open-endedness arises from evolution itself, rather than being a pre-existing condition of the entire evolutionary system. This perspective, championed by Patee and Sayama,~\cite{pattee2019evolved} emphasizes how organisms can create their own mechanisms, such as language, that make late evolution more open-ended. Thus, we must first discern what universal conditions or mechanisms contribute to the emergence of evolutionary open-endedness.

Using a sufficiently general and formal approach to circumscribe any resource-bounded discrete system, it has been shown that state-dependence is a universal mechanism to achieve OEE.~\cite{adams2017formal} However, as the authors recognized, they did not explore self-referential mechanisms between organisms and their environment, which they warned could have enriched their experimental findings (p. 13-14). As we have shown in previous research using the power and generality of category theory to encompass all kinds of mathematical structures (and formal or computational models), semantic closure---the self-referential mechanism where an organism's own actions and interpretations create the context for future actions---is crucial for robust self-replication, niche construction, and evolved open-endedness in any system.~\cite{10.1098/rsif.2025.0784}

In particular, in the case of non-cyclical discrete dynamical systems where there are an infinite number of possible states and each state must contain all the information needed to compute successive states, it has been formally proved undecidability to be a requirement for OEE,~\cite{hernandez2018undecidability} showing that adaptation itself is a generalization of the halting problem.~\cite{turing1936computable} The existence of adapted behaviors whose appearance is formally undecidable reframes evolution as not merely a search through a possibility space, but as an ongoing construction of uncomputable futures.~\cite{kauffman2021world} Therefore, any mechanism associated to the impossibility of constructing a single algorithm that always leads to a correct yes-or-no answer is a potential candidate for promoting open-endedness.

Using Random Boolean Networks (RBNs), we build on earlier work that treated changes in the attractor landscape as a proxy for evolvability. In particular, a selection criteria that reward networks which both conserve existing attractors and generate new ones has been introduced, showing that this conservation-innovation trade-off can drive the evolution of critical dynamics.~\cite{TorresSosa2012} Here we repurpose this attractor-level perspective as an explicit, substrate agnostic metric of open-endedness that is applicable to any dynamical system with discrete state space. Recognizing that open-endedness is inherently a product of evolutionary processes, we further explore the underlying mechanisms driving this phenomenon. Particularly, we focus our attention on non-classical logics: formal systems that modify, extend, or deviate from the foundational principles of classical logic by rejecting or altering concepts such as the law of excluded middle (a statement is either true or false) or bivalence (only two truth values exist).~\cite{priest2008introduction} These alternative frameworks are necessary for modeling situations where classical logic is insufficient, such as dealing with incomplete information, statements that are neither true nor false, paradoxical information, and alternative kinds of knowledge and reasoning. In this way, we investigate the role of undecidability in its different facets to promote open-endedness in a biological model. 

From a systems-biology standpoint, the practical value of $\Omega$ is that it can be computed for any existing discrete state space model, not only for random ensembles. Given a curated Boolean regulatory circuit, one can simulate long trajectories under experimentally motivated sources of variability (e.g., asynchronous update timing, fluctuating inputs, or context switches) and compute $\Omega$ over successive windows. A high $\Omega$ indicates that the model repeatedly visits multiple long-lived expression programs, rather than merely flickering among transient states or collapsing to a single fate. This offers a compact, model-agnostic summary of phenotypic plasticity that can be used to compare alternative network reconstructions, to quantify how mutations or interventions reshape the accessible repertoire of stable phenotypes, and to guide the design of synthetic circuits intended to support rich but structured switching.

This work has two distinct aims that we keep conceptually separate. First, we define and motivate $\Omega$ as a finite-horizon, recurrence-based diagnostic of sustained novelty in discrete dynamical systems, and we specify exactly how it is computed from empirical recurrence episodes (including stochastic dynamics). Second, we use $\Omega$ as an analysis lens for comparing classes of \emph{logic-inspired update mechanisms} (context switching, conditional gating, controlled contradictions, and correlated branching) within a common Random Boolean Network testbed.

To make this separation explicit, our validation proceeds in four steps. (i) We establish analytic sanity checks showing that $\Omega$ vanishes for single-attractor deterministic dynamics and also for “pure noise” dynamics with bounded dwell times (see Methods). (ii) We provide an operational definition of recurrence episodes (cycle discovery, dwell time, escape) that makes clear how chained cycles and non-Boolean intermediate tokens are treated. (iii) We quantify horizon sensitivity and size effects by computing $\Omega$ across multiple observation windows and including a small-$N$ control (Supplementary Figure 22). (iv) We validate portability beyond random RBN ensembles by evaluating $\Omega$ on external benchmarks (Elementary Cellular Automata and curated Boolean GRN models), and we compare $\Omega$ to standard trajectory-level baselines (unique-state fraction, node entropy, and a compression proxy) together with a complement-symmetry control (Supplementary Figure 23, Supplementary Figure 24, and Supplementary Figure 25). Together these checks support interpreting $\Omega$ as a finite-time signature of recurrence-weighted sustained novelty, rather than as a strict $T\to\infty$ attractor decomposition for stochastic dynamics.

This interpretation is intentionally modest. The empirical and biological validity of any OEE indicator depends on the scale at which the evolving system is represented, the variables included in the state description, and the phenotype map used to interpret recurrent states. Therefore, $\Omega$ is best understood as a candidate diagnostic for one aspect of OEE---persistent/recurrent novelty---rather than as a stand-alone criterion for deciding whether a biological system is open-ended in the full evolutionary sense. In particular, because the present implementation fixes the node set and state space in advance, it cannot by itself capture evolutionary events in which the relevant variables, system boundary, or space of possible phenotypes changes over time. Still, $\Omega$ can be measured in cases where the state space is not fixed, since it focuses exclusively on the dynamics of the state space, regardless of whether it is fixed/predefined or not.

In the Results section, we show and examine our findings considering both homogeneous and heterogeneous RBNs, exploring whether structural and temporal heterogeneity also serves as key drivers of open-endedness. Our experiments reveal strategic insights into how systems can maximize their open-endedness based on their inherent conditions and properties. In the Methods section, we detail our methodology, emphasizing the relevance of Random Boolean Networks (RBNs) and non-classical logics in biological modeling, while also introducing our open-endedness metric. In the Discussion section, we contrast our contribution with other approaches to quantify novelty, formulate a generalization of our metric to encompass systems with continuous state space, and propose avenues for future research.

\section{Results} 
We evaluated open-endedness ($\Omega$, see Methods) across classical Random Boolean Networks (RBNs) and four non-Boolean extensions (Paraconsistent, Modal, Quantum-inspired, and Annealed Rule Mutation (ARM)) together with Probabilistic Boolean Networks (PBNs). All of these mechanisms make the dynamics stochastic. In these cases, we do not claim to compute invariant attractors in the strict dynamical-systems sense. Instead, we measure empirical recurrence episodes within a finite horizon using the operational detector described in Methods. Consequently, $\Omega(T)$ should be interpreted as a finite-time recurrence-based signature of sustained novelty, not as a canonical attractor decomposition of the underlying Markov process. Because the state space scales as $2^N$, very long cycles (or extremely slow recurrences) may not be observed within $T=10^6$ steps at $N=100$. Our reported $\Omega(T)$ is therefore a finite-horizon proxy that captures recurrence structure that is actually realized within the observation window. Supplementary Figure 1 and Supplementary Figure 2 report horizon sensitivity checks. The analysis pipeline proceeded as follows.

\begin{enumerate}
    \item First, for each average connectivity $K$ on a fixed grid ($K \in [1.1,4.5]$ in steps of $0.2$), we generated $1,000$ independent RBNs of $N=100$ nodes with random initial states and simulated each network for $T=10^6$ time steps. The full parameter sweeps used to select the settings reported below, as well as sensitivity checks, are provided in from the Supplementary Figure 1 to the Supplementary Figure 25, together with the Supplementary Table 1 and Supplementary Table 2. 
    \item Second, for each base RBN we produced transformed variants by applying the alternative mechanisms to the same structure (see Methods). We then proceeded to run these transformed RBNs under the same conditions.
    \item Third, within each logic we scanned a discrete parameter grid and, for each parameter combination, computed the curve $K \mapsto \Omega(K)$ by averaging $\Omega$ over the $1,000$ networks.
    \item Fourth, to summarize a logic’s overall performance we used the \emph{area under the $\Omega$-$K$ curve} (numerical trapezoidal rule) as a scalar objective and selected, per logic, the parameter combination that maximized this area.
    \item Finally, to compare logics concisely in the main text we plot only these \emph{best} curves for the homogeneous and heterogeneous cases, respectively; all intermediate sweeps and additional diagnostics appear from the Supplementary Figure 3 to the Supplementary Figure 21, the Supplementary Table 1, and the Supplementary Table 2.
\end{enumerate}

We distinguish two architectural regimes. In the \emph{homogeneous} case, networks have structural, temporal, and functional homogeneity; in practice we use a Poisson in-degree distribution (fixed mean $K$), synchronous updating, and a constant bias of $0.5$. In the \emph{heterogeneous} case, we introduce structural and temporal heterogeneity; in practice we use an Exponential in-degree distribution (same mean $K$) and a non-deterministic asynchronous update scheme. In both regimes, the classical baseline is a deterministic BN; PBNs add context switching across deterministic constituents; ARM flips LUT outputs at read-time with probability $\mu$; Paraconsistent logic injects locally contradictory table entries with probability $c$; Modal logic rewrites entries to modal tokens according to $(a,p_p,p_n)$ (accessibility degree and the probabilities of \text{possible}/\text{necessary}); and quantum-inspired mechanisms introduce \text{superposed} outputs with probability $sp$ and pairs up to $e$ pair-coupled nodes (see Methods).

\begin{figure*}[!ht]
  \centering
  \includegraphics[width=\textwidth]{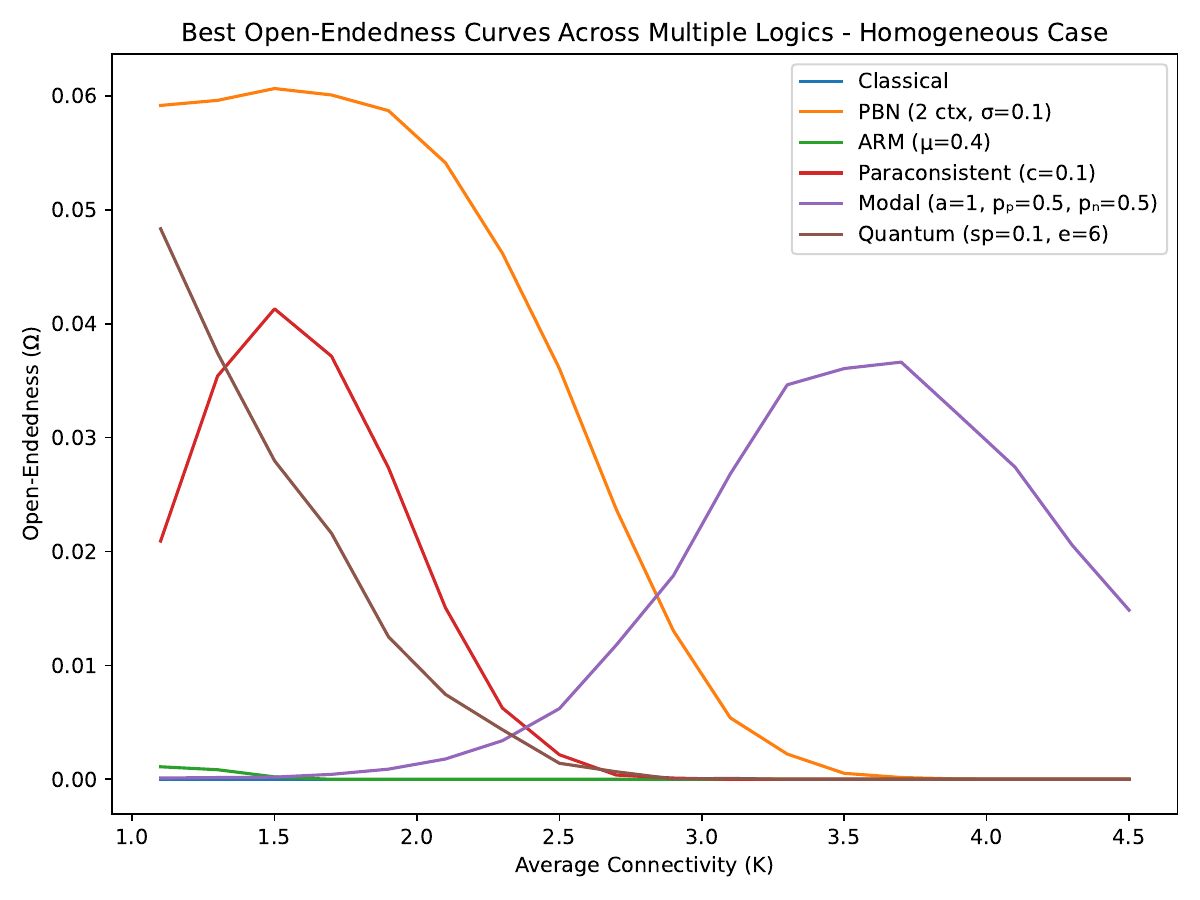}
  \caption{\textbf{Best open-endedness curves across mechanisms - Homogeneous Case.}
  Each curve reports the across-network mean $\Omega(K)$ for the parameter combination within a logic that maximized the area under the curve (AUC) on the $K$ grid ($1.1$ to $4.5$ in steps of $0.2$). Homogeneity comprises Poisson in-degree, synchronous updates, and fixed bias ($0.5$). For every $K$, we averaged over $1,000$ independently generated networks of $N=100$ nodes; each simulation ran for $T=10^6$ time steps from a random initial state. The legend shows the selected (AUC-maximizing) parameters: \emph{Classical} (deterministic baseline); \emph{PBN} “(2 ctx, $\sigma=0.1$)” indicates two deterministic contexts with switching probability $\sigma=0.1$ per epoch; \emph{ARM} “$(\mu=0.4)$” flips each LUT output with probability $0.4$ when read; \emph{Paraconsistent} “$(c=0.1)$” replaces LUT entries by contradictory tokens with probability $0.1$; \emph{Modal} “$(a=1, p_p=0.5, p_n=0.5)$” uses accessibility degree $a=1$, with probabilities $p_p=0.5$ and $p_n=0.5$ of writing \texttt{possible} and \texttt{necessary}, respectively; \emph{Quantum-inspired} “$(sp=0.1, e=6)$” writes \texttt{superposed} with probability $sp=0.1$ and applies paired-state coupling of size $e=6$ (a logic-inspired correlation mechanism; not unitary quantum dynamics). Note that the \emph{Classical} curve collapses to zero, so it is not visible in the plot. Shaded confidence bands are omitted here to emphasize the comparative shapes; full parameter sweeps are provided in the Supplementary Figure 3, Supplementary Figure 4, Supplementary Figure 7, Supplementary Figure 8, Supplementary Figure 11, Supplementary Figure 12, Supplementary Figure 15, Supplementary Figure 16, Supplementary Figure 17, Supplementary Figure 20, and the Supplementary Table 1.}
  \label{fig:homo}
\end{figure*}

Figures~\ref{fig:homo} and \ref{fig:hetero} report, for each regime, the single best curve per logic under the selection rule above. Legends list the winning parameter settings for each mechanism; for consistency we abbreviate parameters as: “ctx” = number of PBN contexts, $\sigma$ = PBN switching probability per epoch, $\mu$ = ARM mutation probability, $c$ = paraconsistent contradiction probability, $a$ = modal accessibility degree, $p_p$/$p_n$ = modal probabilities of writing \text{possible}/\text{necessary}, $sp$ = superposition probability, and $e$ = number of pair-coupled pairs. Axes are shared across panels: the horizontal axis is average connectivity $K$; the vertical axis is our open-endedness score $\Omega$ (see Methods). Now that we have documented the results obtained, let's interpret them.

In homogeneous networks, the PBN transformation achieves the largest $\Omega$ at low-medium connectivities, whereas Modal logic dominates at high connectivities (Fig.~\ref{fig:homo}). The PBN outcome is consistent with the mechanism: piecewise deterministic motion inside a context, punctuated by rare switches, tends to string together long residence times in distinct attractor families, inflating the residence-time-weighted cycle contribution in $\Omega$. Biologically, this mirrors bet-hedging via phenotype switching under resource-scarce conditions: exactly where sparse coupling (low $K$) limits within-context exploration, so context changes do the heavy lifting.~\cite{acar2008stochastic} 
By contrast, when coupling is dense (large $K$), Modal gating becomes advantageous. “Necessary” constraints act as context-dependent canalizing conditions that suppress runaway sensitivity, while “possible” gates open alternative routes when local context permits; together they tame chaotic overshoot and create metastable corridors that repeatedly access distinct limit cycles. As a practical rule: in synchronous, homogeneous systems with limited connectivity, stochastic program switching (PBN) yields the highest open-endedness; as connectivity grows, context-dependent gating (Modal) overtakes by structuring exploration without freezing it.

Moving to heterogeneous networks, the picture changes qualitatively (Fig.~\ref{fig:hetero}). The PBN curve collapses toward zero over the entire $K$ range, while Modal again excels at higher $K$, and Paraconsistent and Quantum-inspired mechanisms show localized gains at low-intermediate $K$. This collapse is not due to a change in the \emph{number} of point attractors, those are determined by the rule tables and are independent of the updating scheme,~\cite{Gershenson2004b} but rather to how our stochastic asynchronous updates interact with structural heterogeneity. In our simulations we use CUBEWALKERS' \texttt{asynchronous\_set} update scheme, which randomly selects a set of nodes to be updated at each time step, so that the dynamics along a single trajectory are effectively noisy and long synchronous cycles are rarely realized. Combined with an Exponential in-degree distribution, this drives most initial conditions quickly into small recurrent classes (often fixed points or very short cycles) that have large basins of attraction.

Within each PBN context, trajectories therefore spend almost all their time in short attractors with cycle lengths $k_j$ of order one; context switches mostly move the system between the basins of these same few attractors, rather than uncovering new long cycles. As a result, the summed residence-cycle contribution $\sum_j d_j k_j$ grows at most linearly with $T$, so $\Omega \rightarrow 0$ despite ongoing context changes. Modal logic is the exception because its necessity/possibility tests provide conditional activation that intermittently re-opens pathways even under asynchronous settling; hence its large spikes at high $K$. Paraconsistent logic yields a broad shoulder around $K \approx 2.3$: introducing controlled local contradictions near the order-critical boundary sustains exploratory dynamics without exploding to noise, in line with the known criticality-extending effects of heterogeneity.~\cite{sanchez2023heterogeneity, lopez2023temporal} Quantum-inspired rules perform best at very low $K$: superposition tokens plus a few long-range paired-state couplings furnish correlated branching when wiring is sparse, boosting early-time exploration before asynchronous contraction dominates. The strategy that follows is therefore different from the homogeneous case: under heterogeneity/asynchrony, prefer Modal gating at high connectivity and use Paraconsistent or Quantum-inspired rules to lift $\Omega$ in sparse networks.

Some patterns are trivial yet worth stating explicitly. In both regimes, the Classical curve sinks over the entire $K$ range because a deterministic network visiting a single attractor contributes $\Omega\rightarrow 0$ for large enough values of $T$ (see Supplementary Figure 1 and Supplementary Figure 2); $\Omega$ rewards persistent traversal of multiple cyclic phenotypes, not mere residence in one. The consistently poor performance of ARM is also instructive: per-read truth-table mutation anneals away stable local structure, pushing dynamics toward effectively memoryless transitions; this drives down the product of cycle length and dwell time and is qualitatively consistent with the classic annealed-approximation intuition that high flip rates erase organized attractor geometry.~\cite{derrida1986random} In short, ARM provides plasticity but not the kind of structured, state-dependent plasticity that $\Omega$ rewards. 

In the next section, we will explore how our metric is related to previous approaches, showing how it is a complementary measure of novelty. As we shall see, our approach focuses specifically on discrete state spaces, leaving aside systems with continuous (or hybrid) state spaces. Thus, additionally, we sketch a principled extension of our metric that preserves its core properties while making it applicable beyond purely discrete models. 

\section{Discussion}
Our starting point is the formal program that sought system-independent definitions of “unbounded evolution” (UE) and “innovation” in discrete dynamical systems.~\cite{adams2017formal} That work emphasizes unboundedness and genuinely novel state production as the core signatures of OEE, abstracted away from any particular substrate or encoding. Our metric is more limited and more operational: we introduce $\Omega(T)$ as a computable finite-horizon diagnostic of recurrence-weighted novelty in discrete state spaces. Rather than claiming that $\Omega$ measures OEE in its entirety, we use it to ask whether a given model trajectory repeatedly discovers and persists in distinct recurrent dynamical phenotypes within a specified observation window. This makes $\Omega$ useful for comparing controlled model classes, but its interpretation remains conditional on the chosen state variables, phenotype map, system boundary, and timescale.

Compared to the \emph{Evolutionary Activity Statistics} (EAS) tradition and the Bedau-Packard classification of long-term dynamics,~\cite{bedau1997comparison, bedau1998classification} our approach dispenses with component-level activity counters and the accompanying choices of neutral baselines. The Bedau program classifies systems via diversity and activity time series, a lens that has been productively used but depends on how components and activity are defined and normalized across domains. Our metric instead interrogates the reachable state space itself---counting distinct phenotypic outcomes realized within a moving window---and therefore avoids reliance on a priori component partitions or domain-specific neutral models while still distinguishing regimes with sustained novelty from those that saturate. In this way it is compatible with, but orthogonal to, EAS-style summaries.

The MODES toolbox reframed measurement around four high-level “hallmarks” (change, novelty, complexity, and ecological potential) offering portable algorithms and inviting lower-level mechanistic metrics to complement them.~\cite{dolson2019modes} Our metric precisely fills that niche: it is a low-level, substrate-agnostic counter of realized phenotypic novelty that can be dropped into any discrete-state model (including RBN variants) and read alongside MODES’ hallmarks. In particular, MODES highlights continual novelty as a central hallmark; our sliding-window $\Omega$ makes this hallmark numerically explicit in systems where “phenotype” is tied to attractors or coarse-grained dynamical outcomes, and it can be summarized jointly with MODES’ other hallmarks to separate sustained novelty from transient bursts. 

Our emphasis on ongoing novelty also dovetails with the algorithmic-information view that relates strong forms of OEE to undecidability. It has been shown that, for a broad class of sophistication-like complexity measures, decidability imposes absolute limits on stable complexity growth and that strong OEE entails undecidability/irreducibility of adapted states.~\cite{hernandez2018undecidability} While our $\Omega$ deliberately does not attempt to estimate sophistication~\cite{kolmogorov1965three} or logical depth,~\cite{bennett1988logical} it is consistent with that result: in regimes where undecidable structure drives unbounded production of phenotypic outcomes, $\Omega$ will continue to rise with the number of distinct attractors realized; in decidable regimes, $\Omega$ saturates. Thus, our metric provides a computable proxy for detecting the phenomenology that undecidability results lead us to expect, without invoking heavy algorithmic-complexity estimators whose incapability of distinguishing novelty from noise has been shown.~\cite{posobin2018random} 

As shown in the Supplementary Figure 23, Supplementary Figure 24, and Supplementary Figure 25, the external benchmarks and baseline comparisons reinforce
the idea that $\Omega$ is a recurrence-weighted persistence diagnostic rather than a raw novelty or
entropy score. In the ECA panel (Supplementary Figure 23), fixed-point rules yield $\Omega(T)\approx 1/T$
(rapid settling), while Rule 54 attains the largest average value, consistent with its well-known
Class-IV localized-structure dynamics on finite periodic lattices. Interestingly, additive Rules 90 and
150 also score highly because finite-ring linear recurrences can generate exact long-period returns,
whereas Rule 110, although also Class-IV, yields smaller $\Omega$ under our random initial condition
protocol because exact whole-lattice recurrences are not reached as efficiently within the finite window.
A fuller explanation of how Wolfram class, additivity, lattice size, and initial-condition ensemble
jointly determine $\Omega$ lies beyond the scope of this paper. 

Baseline proxy comparisons (Supplementary Figure 24) further show that $\Omega$ is not reducible to unique-state growth, entropy, or compressibility alone: high novelty/noise can coexist with low $\Omega$, whereas structured recurrence yields higher $\Omega$ even when novelty proxies are moderate. Finally, the complement-quotient control leaves the qualitative conclusions unchanged in the external benchmark analyses, addressing the concern that attractor diversity could be inflated by trivial Boolean relabelings. The curated Boolean GRN benchmark (Supplementary Figure 25) should likewise be read as a scale comparison rather than as a phase-classification exercise: for small deterministic controls, $\Omega$ mainly reflects realized finite-horizon cycle lengths, whereas the biological point is that curated GRNs remain close to the finite-horizon floor.

Within the York workshop terminology, $\Omega(T)$ is most closely related to the category of “ongoing generation of adaptive novelty,” but only after a phenotype map has been specified.~\cite{taylor2016open} The metric does not determine by itself whether the recurrent states are adaptive, meaningful, or evolutionarily consequential; it only quantifies whether distinct recurrent dynamical phenotypes continue to appear and persist. For this reason, $\Omega(T)$ should be read alongside biological interpretation, environmental context, and, when available, fitness or viability information. Our focus on novelty avoids conflating open-endedness with monotone complexity growth, in line with recent arguments indicating that complexity increases are contingent, often nonadaptive, and concentrated in a few lineages, with many trends pointing in the opposite direction.~\cite{lynch2025complexity}

Updating the York classification, the Tokyo taxonomy distinguishes several forms of OEE, including the ongoing generation of new entities/interactions, the evolution of evolvability, major transitions, and semantic evolution.~\cite{packard2019overview} $\Omega(T)$ is most naturally aligned with the first of these when attractors or coarse-grained recurrent behaviors are treated as phenotypic entities. It may also provide a useful auxiliary signal for the evolution of evolvability when the rate or repertoire of recurrent phenotypes changes across evolutionary windows. However, $\Omega(T)$ alone cannot establish higher-level forms of OEE that involve changing organizational levels, new interpretive relations, or semantic roles. Those cases require an explicit account of how the state variables, phenotype map, and system boundary themselves change through evolution. In short, $\Omega$ equips us with a portable finite-time detector for the axis of “ongoing generation” that both the York and Tokyo taxonomies place at the center of OEE. 

A central limitation of the present framework is that $\Omega(T)$ is computed after specifying a state space, node set, update rule, and observation window. This is appropriate for controlled discrete dynamical models, including RBNs, Boolean GRNs, and cellular automata, but it does not by itself capture evolutionary processes in which the relevant variables, system boundary, or space of possible phenotypes changes over time. In biological evolution, new genes, regulatory modules, ecological interactions, developmental stages, and levels of individuality may become relevant as evolution proceeds.~\cite{yang2026scaling} Such changes are precisely among the phenomena that make OEE difficult to formalize.~\cite{kauffman2000investigations} Therefore, $\Omega(T)$ should be understood as a diagnostic applied to a chosen representation of a system, not as a complete representation of evolution itself.

The value of $\Omega(T)$ can also depend on hidden variables and partial observability. If unobserved variables distinguish states that appear identical in the measured projection, a trajectory may look artificially recurrent. Conversely, if measured variables omit the coordinates on which recurrence occurs, a periodic or quasi-periodic system may appear non-recurrent. Similar issues arise from coarse-graining: overly coarse descriptions can collapse genuinely distinct phenotypes into one recurrent class, whereas overly fine descriptions can convert structured recurrence into apparent noise.~\cite{pessoa2024beyond} For this reason, applications of $\Omega$ to empirical or reconstructed biological systems should report the chosen state variables and coarse-graining explicitly, and, when possible, evaluate robustness across multiple biologically meaningful projections or partitions, as suggested below in the continuous extension of our metric.

These limitations do not invalidate $\Omega$ as a finite-horizon recurrence diagnostic for genuine novelty. Rather, they delimit what it can claim. The score is most informative when the state representation is fixed by a clear modeling question---for example, whether a Boolean regulatory circuit supports multiple recurrent expression programs under a specified perturbation regime. It is less appropriate as a stand-alone criterion for deciding whether an evolving biosphere, ecosystem, or lineage is open-ended in the full evolutionary sense. Capturing those broader cases would require a higher-level framework in which the state space itself can expand, split, or be redefined as new variables and organizational levels become evolutionarily relevant. This is, in fact, another motivation for extending our metric to continuous or hybrid spaces.

To generalize $\Omega$ to continuous (or hybrid) state spaces, let $(X,d)$ be a compact metric space and $\Phi_t$ a measurable flow (or map) on $X$. Choose (i) a finite measurable partition $\Pi=\{C_1,\dots,C_Q\}$ of $X$ that represents the phenotypic coarse-graining of interest and (ii) a tolerance $\varepsilon>0$ that encodes measurement noise or resolution. Given a trajectory $x(t)$, form the symbolic stream $s(t)\in\{1,\dots,Q\}$ by setting $s(t)=q$ whenever $x(t)\in C_q$. We say that the dynamics enters an \emph{$(\varepsilon,\Pi)$-attractor episode} $A_j^{\varepsilon,\Pi}$ at time $t_j^{\mathrm{in}}$ if thereafter the orbit remains in a trapping region $\mathcal{T}\subseteq X$ and returns within distance $\varepsilon$ of some previously visited state in $\mathcal{T}$ before time $t_j^{\mathrm{out}}$; the \emph{residence time} is $d_j^{\varepsilon,\Pi}=t_j^{\mathrm{out}}-t_j^{\mathrm{in}}$. For that episode define the \emph{symbolic cycle-length}

\begin{equation}
    K_j^{\varepsilon,\Pi} \;=\; \bigl|\{\, s(t)\;:\; t\in [t_j^{\mathrm{in}},\,t_j^{\mathrm{out}})\,\}\bigr|,
\end{equation}

i.e., the number of distinct partition elements visited before the $\varepsilon$-return (this coincides with the usual cycle-length in a discrete BN when $\Pi$ separates individual states and $\varepsilon\!\to\!0$). Denoting by $m(T)$ the number of such episodes discovered up to time $T$, we set 

\begin{equation}
\Omega_{\varepsilon,\Pi}(T)\;=\;\frac{1}{T^2}\sum_{j=1}^{m(T)} d_j^{\varepsilon,\Pi}\, K_j^{\varepsilon,\Pi},
\end{equation}

defining open-endedness for continuous state spaces as

\begin{equation}
\Omega_{\varepsilon,\Pi}\;=\;\limsup_{T\to\infty}\Omega_{\varepsilon,\Pi}(T).
\label{eq:Omega-cont}
\end{equation}

This extension preserves the key behavior of our discrete metric: (a) if the system converges to a single fixed point or limit cycle contained within one recurrent class of $\Pi$, then $\Omega_{\varepsilon,\Pi}=0$ because there is only one episode and $d_1^{\varepsilon,\Pi}=O(T)$ while $K_1^{\varepsilon,\Pi}$ is $O(1)$; (b) if the number of distinct $(\varepsilon,\Pi)$-attractor episodes diverges, then $\Omega_{\varepsilon,\Pi}$ diverges as well. Moreover, by the same Cauchy-Schwarz argument used in the discrete case we obtain a resolution-dependent upper bound

\begin{equation}
\Omega_{\varepsilon,\Pi}
\;\le\;
\limsup_{T\to\infty}\frac{1}{T^2}\Bigl(\sum_{j=1}^{m(T)} d_j^{\varepsilon,\Pi}K_j^{\varepsilon,\Pi}\Bigr)^2
\;\le\;
m(\varepsilon,\Pi)\,|\Pi|^2,
\end{equation}

since $\sum_j d_j^{\varepsilon,\Pi}\le T$ and $(K_j^{\varepsilon,\Pi})^2\le |\Pi|^2$. Thus the \emph{same monotone relationship with the number of recurrent phenotypes} holds as before, and when $\Pi$ refines to singletons and $\varepsilon\to0$, \eqref{eq:Omega-cont} recovers our original definition (see Methods). Importantly, Eq.~\eqref{eq:Omega-cont} lets us assess open-endedness in any continuous (or hybrid) state space system by working with a biologically meaningful coarse-graining $\Pi$ (e.g., thresholds on gene expression) and a resolution $\varepsilon$ reflecting experimental noise. 

Two challenges naturally follow. First, $\Pi$ can be fixed from prior biological semantics (cell fates, pathway activities) or learned as a Markov partition-like discretization that preserves long-time statistics. In continuous systems, exact revisits of a microstate have probability zero, so all recurrences are defined relative to a finite resolution $\varepsilon$ and partition $\Pi$. For a given $(\varepsilon,\Pi)$ it may happen that no coarse-grained state is ever revisited along a finite trajectory, in which case $\Omega_{\varepsilon,\Pi}=0$: at that scale the dynamics look like pure exploration with no recurrent patterns. This does not preclude open-endedness at finer or more structured resolutions. As we refine the partition or decrease $\varepsilon$, the same trajectory can reveal recurrent structure (e.g., visits to a fractal subset within each cell), and by construction $\Omega_{\varepsilon',\Pi'} \ge
\Omega_{\varepsilon,\Pi}$ for $\varepsilon'\le\varepsilon$ and $\Pi'\succ\Pi$. A principled program is therefore to study $\Omega_{\varepsilon,\Pi}$ under partition refinement, reporting curves of $\Omega_{\varepsilon,\Pi}$ against $|\Pi|$ and $\varepsilon$, looking for regimes where $\Omega_{\varepsilon,\Pi}$ grows and then stabilizes. This multi-scale analysis plays a role analogous to box-counting methods for fractal dimension, and avoids committing to a single “correct” coarse-graining: overly coarse partitions indeed wash out novelty, whereas overly fine ones see only noise.

Second, for long trajectories, exact recurrence detection is costly. In continuous state spaces many trajectories will cross separatrices, escape to infinity, or otherwise hit “points of no return”, so exact state revisits are not expected globally. In practice, we would first identify \emph{coarse-grained recurrent classes} (quasi-attractors) by looking for approximate returns to the same cell of a finite-resolution partition. Our GPU streaming/extractor could be extended to use fast $\varepsilon$-recurrence queries (e.g., locality-sensitive hashing of Poincaré samples or recurrence-plot-based triggers), guaranteeing $O(1)$ amortized checks per step. Trajectories that never return to a cell at the chosen resolution simply do not generate a recurrent class there, and thus contribute $\Omega_{\varepsilon,\Pi}=0$ at that scale. Once such coarse-grained recurrent classes have been detected, their residence times $d_j^{\varepsilon,\Pi}$ become random variables. We then could estimate $\Omega_{\varepsilon,\Pi}$ by ergodic averages over long runs, relating it to return-time statistics \emph{within} each coarse-grained recurrent class (i.e., quasi-attractor), rather than assuming global recurrence of the full system.

\begin{figure*}[!ht]
  \centering
  \includegraphics[width=\textwidth]{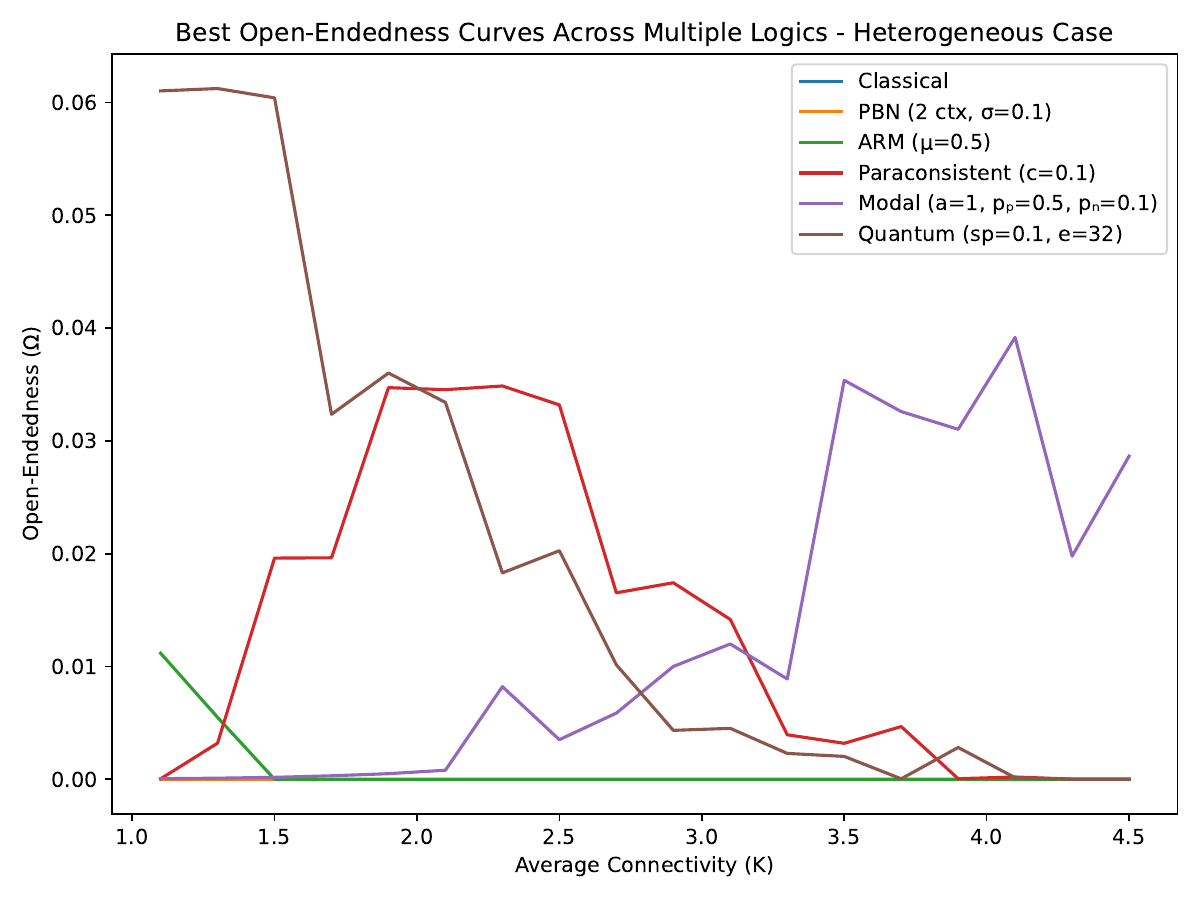}
  \caption{\textbf{Best open-endedness curves across mechanisms - Heterogeneous Case.}
  Same plotting and selection protocol as Fig.~\ref{fig:homo}, but for structurally and temporally heterogeneous networks (Exponential in-degree; stochastic asynchronous updates). For each $K$, results average $1,000$ networks of $N=100$ nodes simulated for $T=10^6$ steps. The AUC-maximizing parameter sets (legend) are: \emph{Classical} (deterministic baseline); \emph{PBN} “(2 ctx, $\sigma=0.1$)”; \emph{ARM} “$(\mu=0.4)$”; \emph{Paraconsistent} “$(c=0.1)$”; \emph{Modal} “$(a=1, p_p=0.5, p_n=0.1)$”; and \emph{Quantum-inspired} “$(sp=0.1, e=32)$”. As in Fig.~\ref{fig:homo}, $\Omega$ values are residence-time weighted cycle-lengths normalized by $T^2$ (see Methods). Note that both \emph{Classical} and \emph{PBN} curves collapse to zero, so they are not visible in the plot. All intermediate sweeps, robustness checks (e.g., alternative $K$ grids and $T$ values), and per-logic ablation plots appear in the Supplementary Figure~5, Supplementary Figure~6, Supplementary Figure~9, Supplementary Figure~10, Supplementary Figure~13, Supplementary Figure~14, Supplementary Figure~18, Supplementary Figure~19, Supplementary Figure~21, and the Supplementary Table~2.}
  \label{fig:hetero}
\end{figure*}

Overall, the experiments here presented use random network ensembles and stylized state-dependent operators as a controlled testbed. We do not claim that specific non-classical logics are literally implemented in gene regulation. Rather, these operators serve as mechanistic motifs for injecting context dependence and structured uncertainty into discrete regulatory dynamics. In empirical applications, $\Omega$ should be computed on curated GRN models with biologically justified update/noise assumptions and interpreted relative to the chosen phenotype coarse-graining and finite simulation horizon. Even so, our results allow us to propose the following actionable design principles for maximizing open-endedness:

\begin{enumerate}
    \item In homogeneous networks with synchronous updates and low connectivity $K$, our simulations show that temporal heterogeneity is the main lever for raising open-endedness. When context switches are rare, each context relaxes into a long-lived attractor, and the occasional switch effectively concatenates long-residence segments from different attractor families along a single trajectory, yielding large values of $\Omega$. In this regime both the wiring and the rule distribution are homogeneous; the only source of heterogeneity is the stochastic context schedule. When we later introduce structural heterogeneity (Fig.~\ref{fig:hetero}), the same context-switching protocol no longer boosts $\Omega$ in the same way, indicating that structural and temporal heterogeneity interact non-trivially rather than one simply dominating the other. 
    \item Dense high-$K$ networks in the homogeneous, synchronously updated ensembles show their strongest open-endedness under Modal logic. Here, context-dependent necessity/possibility acts as a filter that prunes most chaotic branches while still keeping alternative pathways available, so trajectories remain in long, structured attractors rather than collapsing immediately to fixed points. When we later introduce structural heterogeneity and non-deterministic asynchronous updates (Fig.~\ref{fig:hetero}), the same qualitative advantage of Modal logic at high $K$ persists, but the overall magnitude of $\Omega$ is reduced by the rapid contraction induced by asynchronous settling.
    \item For heterogeneous/asynchronous systems: Modal at high $K$ and Paraconsistent or Quantum-inspired mechanism at low-intermediate $K$ maximize open-endedness. 
    \item Avoid high mutation rates when the goal is sustained novelty: ARM’s indiscriminate flips wash out the attractor geometry that $\Omega$ detects.
\end{enumerate}

We note two cross-checks that bolster these interpretations. First, the high-$\Omega$ zones for Paraconsistent and Modal variants align with regions where heterogeneity is known to extend or reshape the critical regime in Boolean networks.~\cite{lopez2023temporal} Second, the PBN advantage under homogeneity but not heterogeneity is consistent with the literature on update schemes: when all nodes are updated synchronously, the dynamics follow a fixed deterministic map that can sustain rich cyclic structure, whereas \emph{stochastic} asynchronous updates tend to  break long cycles and contract dynamics into small recurrent classes.~\cite{Gershenson2004b} In our heterogeneous simulations we combine an Exponential in-degree distribution with CUBEWALKERS’ non-deterministic \texttt{asynchronous\_set} scheme, so it is this added randomness in the update order, rather than asynchronicity \emph{per se}, that erodes the attractor geometry that $\Omega$ rewards. Deterministic asynchronous schemes with fixed node orderings can, by contrast, preserve or even lengthen cycles, so our claim is restricted to the non-deterministic case we study here.

In the aggregate, Figures~\ref{fig:homo}-\ref{fig:hetero} suggest that state-dependence is necessary but not sufficient; how it is introduced---via context switching, conditional gating, controlled contradictions, or correlated branching---interacts strongly with architectural conditions (connectivity and update heterogeneity) to determine whether open-ended dynamics flourish or fade. This points towards a design principle for engineering OEE: choose the mechanism whose \emph{mode of uncertainty} (what is randomized, when, and where) best complements the system’s wiring and timing. In this sense, our results are analogous to the design of \emph{the right amount} of
antifragility in Boolean networks. It has been shown that the benefit a system derives from perturbations depends on how the level and pattern of external noise are matched to the underlying dynamics, and identified regimes where perturbations improve a satisfaction functional rather than merely degrading performance.~\cite{pineda2019novel} Here we obtain a complementary view: instead of tuning environmental noise for a fixed network, we tune the internal realization of uncertainty for a given architecture, and assess its effect on the long-term production of novel attractors via $\Omega$. 

Our work then suggests several concrete future research threads. It is possible to explore other mechanisms that promote open-endedness, including the composition of the multiple non-Boolean logics we study here. Our code already supports composition mechanisms, so the next step would be to optimize logic parameters (and their schedules) to maximize $\Omega$ subject to resource constraints, turning our pipeline into an automatic “OEE controller” for synthetic circuits. Additionally, because $\Omega$ is intentionally agnostic to meaning, we can pair it with a task- or fitness-based functional to separate “mere novelty” from \emph{adaptive} novelty, retaining the York/Tokyo emphasis on ongoing generation while remaining system-independent. To our knowledge, no measure of semantic open-endedness has been explored in the literature, so generalizing the metric proposed here in that direction would be a valuable contribution.

Since we just came up with a continuous generalization of our metric, we could apply it to continuous GRN models (such as Hill-function ODEs,~\cite{ehsan2018method} piecewise-linear Glass systems,~\cite{edwards2007periodicity} random fuzzy networks, \cite{Zapata2020} etc.) using phenotype-aware partitions and compare with the discrete-RBN results to test which undecidability-adjacent mechanisms (modal gating, controlled contradictions, context switching) remain most effective in the continuous limit. Moreover, we could relate $\Omega_{\varepsilon,\Pi}$ to invariants such as topological entropy~\cite{adler1965topological} or return-time spectra in chaotic attractors~\cite{boev2014poincare} and determine the conditions under which $\lim_{\varepsilon\to0}\lim_{\Pi\to\text{refined}}\Omega_{\varepsilon,\Pi}$ exists and is equal to the discrete-state $\Omega$ for Boolean embeddings. This opens a path to compare, on equal footing, discrete and continuous regulatory schemes across modeling paradigms, and to design interventions that promote sustained novelty under real biophysical constraints. 

\section{Methods}

\subsection{Random Boolean Networks}
Boolean network models provide a widely used coarse-grained description of gene regulatory networks (GRNs) in which each gene is represented as an ``active''/``inactive'' variable and regulation is encoded as a directed interaction graph.~\cite{Kauffman1969, gershenson2004introduction} We use Random Boolean Networks (RBNs) as a controlled ensemble in which wiring, logic, and update timing can be systematically manipulated while retaining the standard biological interpretation of attractors as putative cell fates.

An RBN is specified by $N$ nodes, an in-neighborhood $I_i$ for each node $i$ (with $|I_i|=k_i$), and a Boolean update rule $f_i:\{0,1\}^{k_i}\to\{0,1\}$ (conveniently represented as a lookup table, alias LUT). Writing the network state as $\mathbf{x}(t)=(x_1(t),\ldots,x_N(t))\in\{0,1\}^N$, synchronous dynamics are given by
\begin{equation}
x_i(t+1)=f_i\!\left(\mathbf{x}_{I_i}(t)\right).
\end{equation}
In the classical Kauffman ensemble, regulators and truth tables are sampled at random; in our simulations we draw regulators according to a chosen in-degree distribution and sample each lookup table with bias $p$ (we use $p=0.5$ unless stated otherwise).

To probe architectural effects, we consider two regimes. In the \emph{homogeneous} regime we use a narrow in-degree distribution (Poisson with mean $K$), synchronous updates, and fixed (quenched) truth tables. In the \emph{heterogeneous} regime we introduce structural and temporal heterogeneity by using a broader in-degree distribution (Exponential with the same mean $K$) together with stochastic asynchronous updates implemented with CUBEWALKERS.~\cite{park2023models} This heterogeneity protocol follows previous work on heterogeneity in Boolean networks;~\cite{lopez2023temporal} here we summarize only what is needed to reproduce our simulations and focus on how different state-dependent mechanisms modify the dynamics. Once we have introduced RBNs, let us discuss the biological role of non-classical logics and how we can implement them using RBNs. 

\subsection{Probabilistic Boolean Network (PBN)} 
Stochastic phenotype switching is a well-established adaptation strategy in fluctuating environments, classically demonstrated in many bacterial systems.~\cite{acar2008stochastic} PBNs were introduced precisely to capture rule-based uncertainty and context switching in GRNs, and have since become a standard for modeling, inference, and control of genetic networks.~\cite{shmulevich2010probabilistic} A PBN is a collection of Boolean networks in which a constituent network governs gene activity for a random period of time before another randomly chosen constituent network takes over. This makes PBNs the most direct way to incorporate environmental contingency and regulatory program switching into an otherwise Boolean framework. 

Formally, a PBN consists of a sequence $V = \{x_i\}_{i=1}^n$ of $n$ nodes, where $x_i \in \{0,1\}$, and a sequence $\{\textbf{f}_l\}_{l=1}^m$ of vector-valued functions, defining constituent networks. Each vector-valued function $\textbf{f}_l = ( f_l^{(1)}, f_l^{(2)}, ..., f_l^{(n)})$ determines a constituent network, or context, of the PBN. The function $f_l^{(i)}: \{0,1\}^k \rightarrow \{0,1\}$ is a predictor of gene $i$ whenever network $l$ is selected. At each updating epoch a decision is made whether to switch the constituent network. A constituent network is randomly selected from among all constituent networks according to the selection probability distribution $\{c_l\}_{l=1}^m$ such that $\sum_{l=1}^m c_l = 1$. In our code, we generate $m$ pure-Boolean contexts by resampling lookup tables (LUTs) from a base Boolean Network (BN) and set uniform $c_l$. We then execute long segments in the current context, updating deterministically within that context and only sampling a new context at segment boundaries. This cleanly reproduces the standard PBN semantics used in the gene-regulation literature.~\cite{shmulevich2010probabilistic}

\subsection{Annealed Rule Mutation (ARM)}
GRN logic is not immutable: transcriptional programs can rewire or tune their effective rules across time due to chromatin remodeling,~\cite{filtz2014regulation} transcription factor post-translational modifications,~\cite{swift2017matter} or changing cofactor abundance.~\cite{tomljanovic2025epigenetic} From a modeling standpoint, this is the annealed perspective familiar from RBN theory, where rule tables are treated as stochastic variables across updates, capturing plasticity not represented by fixed (“quenched”) rules. Empirically, modern time-resolved genomics underscores that regulatory interactions are dynamic over minutes-hours, supporting annealed or partially annealed abstractions when studying innovation and adaptability.~\cite{meeussen2024time} The annealed approximation proposed by Derrida~\cite{derrida1986random} remains a standard conceptual scaffold, and recent reviews emphasize explicitly time-dependent GRN modeling.~\cite{marku2023time}

Formally, an ARM network is a BN in which each LUT entry $f_i(\cdot)$ is a \emph{mutable rule} with parameter $p_{\text{ARM}}\in[0,1]$. Concretely, for an input key $u$, the stored value is a record $\theta(u) = \{\text{initial value}\in\{0,1\}, \text{mutation probability} = p_{\text{ARM}}\}$. The evaluator $\eta$ samples the output each time step by flipping the initial bit with probability $p_{\text{ARM}}: \text{Bernoulli}(1-p)\times\text{initial value} + \text{Bernoulli}(p)\times(1-\text{initial value})$. Thus, $x_i(t+1) = \eta\left(\theta\left(x_i(t)\right)\right)$. If $p_{\text{ARM}}=0$, we recover a quenched BN; $p_{\text{ARM}}>0$ produces an annealed mutation rate in the gate outputs. In our code, we flip the stored bit with probability $p_{\text{ARM}}$ each time it is read. In practice, this means our update loop should pass LUT entries before computing the node’s next value, thereby sampling an annealed truth-table at run time. This ARM mechanism is distinct from PBNs (which switch entire contexts); here, individual rules mutate locally and continuously, providing a complementary axis of plasticity.

\subsection{Paraconsistent Logic} 
Cells frequently face conflicting regulatory cues: the same upstream factor can both promote and inhibit a response depending on context, dosage, or timing.~\cite{hart2012design} This is well documented in immune and developmental systems, e.g., IL-2 simultaneously supporting proliferation and death in T cells;~\cite{hart2013utility} such circuits exploit antagonistic arms to achieve robustness, fold-change detection, and pulsatile dynamics, hence called “paradoxical components.”~\cite{hart2014paradoxical} These phenomena motivate logics that tolerate local contradictions rather than exploding to triviality, which is exactly what paraconsistent formalisms provide. \cite{Gershenson1999} Using an inconsistency-tolerant semantics at the rule/local-gate level mirrors how GRNs integrate pleiotropic, noisy, or partially contradictory inputs without collapsing function.

Formally, let $V=\{x_i\}_{i=1}^n$ be nodes with Boolean states $x_i\in\{0,1\}$. Each node $i$ has LUTs $f_i:\{0,1,\times\}^{k}\rightarrow\{0,1,\times\}$, where $\times$ denotes an inconsistent output. In our implementation, inconsistency is annotated: some LUT entries are replaced by pairs $(2, b)$ with contradiction probability $\pi\in[0,1]$. At update time, a resolver $\rho$ maps regulator inputs $u\in\{0,1,(2,b)\}^{n}$ to a determinate value by discarding 2's where possible and returning unanimity if present; otherwise it returns $(2, b^*)$ with $b^*$ the local majority. This the network's global update is $x_i(t+1) = \rho\left(f_i\left(x_i(t)\right)\right)$, which preserves non-explosive behavior under local contradictions. In our code, each LUT entry is independently replaced by a pair as described above with certain probability $\pi$, such that contradictions can persist and propagate until later steps resolve them via majority among incoming signals, closely matching a three-valued, inconsistency-tolerant gate semantics; echoing how paradoxical circuits stabilize responses in biology.

\subsection{Modal Logic} 
Many regulatory claims in biology are modal: a factor may be necessary (must be present) or merely possible/sufficient (can enable) for an outcome, and these claims are inherently contextual (cell type, chromatin state, signaling milieu).~\cite{monteiro2008temporal} The systems biology community routinely encodes such “must/possibly” properties with temporal/modal logics and verifies them on GRN models via model checking, e.g., verifying that “eventually gene A activates B and always C stays off” in stress-response or cell-cycle networks.~\cite{monteiro2014model} This makes modal operators a natural abstraction layer for contextual constraints in GRN dynamics. Representative overviews and applications include model checking for logical regulatory networks and temporal-logic pattern mining in cellular models.~\cite{giacobbe2015model}

Formally, a modalized RBN augments a standard RBN with a Kripke frame $\mathcal{K} = (V, R)$, where $V$ is the set of nodes and $R\subseteq V\times V$ encodes \emph{accessibility} among nodes. Each LUT entry for node $i$ may take values in $\{0,1,\text{possible},(\text{necessary}, v)\}$. At time $t$, let $A_i(t) = \{j\in V | (i,j)\in R\}$ be $i$'s accessible set, and let $\textbf{x}_{A_i}(t)$ be their current states. The evaluation map $\mu$ is: $\mu(\text{possible}) = 1$ iff $\exists j\in A_i(t): x_j(t)=1$; $\mu\left((\text{necessary}, v)\right) = v$ iff $\forall j \in A_i(t):x_j(t)=1$, else 0; and $\mu(b) = b$ for $b\in\{0,1\}$. The network update is $x_i(t+1) = \mu(f_i(x_i(t)))$. In our code, we build a Kripke frame as described above and sample an accessibility degree $d$, wiring each node to up to $d$ accessible nodes. It then rewrites LUT outputs by flipping entries to the literal tokens \emph{possible} or \emph{necessary} (the latter carrying a concrete bit) with probabilities $p_{\text{possible}}$ and $p_{\text{necessary}}$. This cleanly imports necessity/possibility semantics into our RBN while keeping the base state space Boolean.

\subsection{Quantum-inspired correlated branching} 
While canonical GRNs are classical, quantum-inspired rules can model contextuality and “superposition-like” uncertainty during decision points (e.g., fate priming).~\cite{cao2020quantum} Moreover, quantum coherence and related effects have been directly observed in biological light-harvesting complexes (e.g., the Fenna-Matthews-Olson bacteriochlorophyll complex), suggesting that non-classical information processing can appear in biology under some conditions---even if not in gene regulation \emph{per se}.~\cite{engel2007evidence, panitchayangkoon2010long} The point here is not that GRNs are quantum, but that quantum-style operators offer a principled way to represent context-dependent mixture/branching and long-range correlations that classical gates do not capture natively.~\cite{franco2021random} We emphasize that our goal is \emph{not} to implement formal quantum logic (e.g., non-distributive propositional lattices~\cite{hartonas2018order}) nor quantum mechanics (e.g., unitary evolution in a Hilbert space~\cite{das2021reproducing}). 

Instead, we introduce a \emph{quantum-inspired} mechanism that captures two qualitative motifs often associated with “quantum-like” behavior in complex systems: (i) stochastic resolution of an underspecified update (a superposition-inspired token that collapses to a Boolean value) and (ii) long-range instantaneous correlation implemented as a \emph{paired-state coupling} between nodes. Concretely, LUT entries may take values in $\{0,1,\texttt{superposed}\}$. At update time, an evaluator $\kappa$ maps \texttt{superposed} to a Bernoulli$(1/2)$ draw in $\{0,1\}$, so the realized network state remains Boolean. Separately, we sample a set of paired couplings $E\subseteq V\times V$ (a set of node pairs). If node $i$ is paired with node $j$, then $x_i(t+1)=x_j(t)$, which acts as a classical proxy for instantaneous correlation. This construction is therefore best understood as a logic-inspired correlated-branching update rule, not as a claim about genuine quantum dynamics. Now we can proceed to introduce our open-endedness metric.

\subsection{A Novel Metric for Quantifying Open-Endedness in Any Discrete State Space} \label{sec:episode-detection}

In the deterministic case, regardless of their initial configuration, all Boolean networks fall into an attractor.~\cite{gershenson2004introduction} Attractors are stable sets of states that represent the long-term behavior of the network. Once a Boolean network reaches an attractor, it will remain within that set of states indefinitely unless perturbed. Attractors can be single states (point attractors) or cycles of multiple states (cycle attractors) and are often linked to biological phenotypes, such as cell types in GRNs.~\cite{Kauffman1969} For any attractor in a Boolean network, we can calculate the number of unique states the network cycles through repeatedly (cycle-length) and the number of time steps the network remains trapped within the attractor (residence-time). In this way, given a RBN with $N$ nodes where $m$ different attractors appear during $T$ time steps, we define our finite-horizon recurrence-weighted novelty score ($\Omega$) as

\begin{equation}
   \label{eq:Omega}
    \Omega = \lim_{T\to\infty}\frac{1}{T^2}\sum_{j=1}^m d_j\,k_j~,
\end{equation}

where $k_j$ and $d_j$ are respectively the cycle-length and residence-time of the attractor $A_j$ discovered within $T$ time-steps. In all experiments of the present manuscript we use an operational finite-horizon recurrence detector that extracts attractor episodes from a single simulated trajectory, rather than attempting a full invariant (at $T\to\infty$) decomposition of the underlying dynamics. Although several mechanisms introduce non-Boolean \emph{rule-level} tokens (e.g., modal \texttt{possible}/\texttt{necessary}, quantum-inspired \texttt{superposed}, etc.), the network state vector that we record at each time step remains Boolean. Rule outputs are evaluated through a semantics map (e.g., modal evaluation, superposition “collapse”, etc.) that produces a Boolean update. Thus, recurrence is always detected on realized Boolean microstates, not on intermediate symbolic labels.

Let $\{x(t)\}_{t=0}^{T}$ be the realized Boolean trajectory. We maintain a hash map $\mathrm{firstSeen}[x]$ that stores the earliest time index at which each state $x$ appeared. At time $t$ we declare a cycle discovery when $x(t)$ matches a previously seen state
$x(\tau)$ with $\tau<t$ \emph{and we are not currently inside an episode}. We then set the cycle-length
$k := t-\tau$ and start a new episode anchored at $\tau$. Once an episode starts, its dwell time $d$ is the number of subsequent time steps the
trajectory continues to spend inside the previously explored recurrent region. Operationally, $d$ increments by $1$ each
step until the trajectory hits a \emph{never-before-seen} state (a state not in $\mathrm{firstSeen}$), which we treat as an escape. While inside an episode, additional returns to already visited states do
\emph{not} create new episodes. In particular, if the trajectory traverses multiple short loops that share states, or returns to a prior
loop after visiting another loop (a “figure-eight” / chained-cycle pattern), this is counted as \emph{one continuous recurrent episode}
rather than multiple independent attractors. 

In this way, a new episode can begin only after at least one genuinely novel state appears. Below is the minimal logic used to extract $(k_j,d_j)$ from a trajectory (notations correspond to the codebase): 

\begin{quote}\small\ttfamily
firstSeen = \{\}; episode = None \\
for t = 0..T-1: \\
\ \ s = hash(x(t)) \\
\ \ if episode is None and s in firstSeen: \\
\ \ \ \ tau = firstSeen[s];\ \ k[tau] = t - tau;\ \ d[tau] = 1;\ \ episode = tau \\
\ \ else if episode is not None: \\
\ \ \ \ d[episode] += 1 \\
\ \ \ \ if s not in firstSeen:\ \ episode = None \\
\ \ if s not in firstSeen:\ \ firstSeen[s] = t
\end{quote}

We report the finite-horizon proxy
$\Omega(T)=\frac{1}{T^2}\sum_j d_j k_j$ computed from the above episode detector. For deterministic systems this coincides with standard cycle detection once the orbit enters a true limit cycle. For stochastic systems, $\Omega(T)$ instead summarizes empirical recurrence structure within a finite horizon, i.e., how often the dynamics returns to previously visited microstates and how long it continues to revisit the explored recurrent region. This is a recurrence-based diagnostic, not a strict Markov-chain attractor decomposition. In intuitive terms, $\Omega(T)$ measures how much of the observed trajectory is spent in recurrent dynamical patterns, weighted by the length of those recurrent cycles. In the deterministic case, when $m=1$, we have a single attractor of length $k_1\in\{1,2,\dots,2^N\}$. If we enter this attractor at time step $t_0$, then

\begin{equation}
    \label{eq:nonOEE}
    \Omega = \lim_{T\to\infty}\frac{1}{T^2} d_1\,k_1 = \lim_{T\to\infty}\frac{1}{T^2}\,(T-t_0)\,k_1
       = 0.
\end{equation}

Thus, according to $\Omega$, a system with a single attractor has zero recurrence-weighted novelty in the asymptotic limit. This is totally coherent to what we have operationalized as a proxy for open-endedness: having a single attractor implies that the system, regardless of its initial state (within a certain basin of attraction), will eventually settle into a specific, predictable long-term behavior represented by that single attractor. It is also instructive to consider the opposite extreme: a maximally random process with lots of novelty but minimal dwell time. Suppose that at each time step the network jumps independently to a randomly chosen configuration in $\{0,1\}^N$. We may treat each visited configuration as a point attractor with $k_j = d_j = 1$. After $T$ steps we have $m(T)=T$ such “attractors”, so

\begin{equation}
  \Omega = \lim_{T\to\infty}\frac{1}{T^2} \sum_{j=1}^T 1\cdot 1 = \lim_{T\to\infty}\frac{T}{T^2} = \lim_{T\to\infty}\frac{1}{T} = 0
\end{equation}

In other words, a process that constantly produces new states but never dwells on any of them receives a vanishing $\Omega$ value: its novelty is not recurrently stabilized. The same conclusion holds for a “slow” or delayed random walk on a finite state space that includes self-loops, as long as the expected residence-time in any configuration remains bounded while $T$ grows. In that case, the numerator $\sum_j d_j k_j$ can grow at most linearly with $T$, so $\Omega(T) = \mathcal{O}(1/T)$ and still vanishes in the limit. For a deterministic map on a finite state space, every trajectory eventually becomes periodic.
Let $\mu$ be the first time at which the trajectory revisits a previous state (the entry time into the eventual cycle), and let $\lambda$ be the corresponding cycle length. For any horizon $T>\mu$, the trajectory spends the remaining $(T-\mu)$ steps on the same cycle, so the operational sum in Eq.~\eqref{eq:Omega} reduces to a single attractor episode with $k=\lambda$ and $d(T)=T-\mu$, yielding

\begin{equation}
\label{eq:Omega_det_scaling_SM}
\Omega(T)=\frac{1}{T^2}\,\lambda\,(T-\mu)=\frac{\lambda}{T}\left(1-\frac{\mu}{T}\right)\;=\;\mathcal{O}\!\left(\frac{1}{T}\right).
\end{equation}

Thus, once the eventual cycle has been reached and identified, $\Omega(T)$ must decay approximately as $1/T$ as the observation window grows. More generally, for any fixed finite state space one has $\sum_j d_j k_j \le (2^N)\,T$, hence $\Omega(T)\le 2^N/T$ and the strict $T\to\infty$ limit vanishes even for stochastic dynamics; the practical role of $\Omega(T)$ is therefore as a \emph{finite-horizon recurrence diagnostic}. In the Supplementary Figure 22, we explicitly check the horizon dependence for both the deterministic RBN baseline and a stochastic PBN mechanism, and include a small-$N$ control to illustrate when the asymptotic $1/T$ regime becomes observable. 

To address external validation, index comparisons, and robustness to trivial relabelings, we evaluated $\Omega$ on (i) a small panel of Elementary Cellular Automata (ECA) rules spanning fixed/periodic/complex regimes, and (ii) curated Boolean gene-regulatory network (GRN) models. In addition, we computed three common trajectory-level proxies on the same runs (unique-state fraction, mean node entropy, and a compression proxy) and recomputed $\Omega$ under the complement quotient that identifies each Boolean state $s$ with its complement $1-s$. These results are reported in Supplementary Figure 23, Supplementary Figure 24, and Supplementary Figure 25.

Taken together, these results show that non-zero $\Omega$ requires not only the continual appearance of distinct attractors, but also that the total time spent in structured, recurrent dynamical patterns grows faster than linearly with $T$. Now let us assume that the dynamics are no longer deterministic or completely random. By construction, there exists an ordered sequence of time steps, $\{t_1^i, t_1^f, t_2^i, t_2^f, \dots, t_m^i, t_m^f\}$, such that $A_j$ appears in $[t_j^i, t_j^f]~~~\forall j\in\{1,\dots,m\}$. This implies the following chain of equalities and inequalities:

\begin{align*}
    \Omega &= \lim_{T\to\infty}\frac{1}{T^2}\sum_{j=1}^m d_j\,k_j \\
    &= \lim_{T\to\infty}\frac{1}{T^2} \sum_{j=1}^m (t_j^f - t_j^i)k_j\\
    &\leq \lim_{T\to\infty}\frac{1}{T^2}\left[\sum_{j=1}^m (t_j^f - t_j^i)k_j\right]^2\\
    &\leq \lim_{T\to\infty}\frac{1}{T^2} \left[\sum_{j=1}^m (t_j^f - t_j^i)^2\right]\left[\sum_{j=1}^mk_j^2\right]\\
    &\leq \lim_{T\to\infty}\frac{1}{T^2} \left[\sum_{j=1}^m (t_j^f - t_j^i)\right]^2 \left[\sum_{j=1}^m \left(2^N\right)^2\right]\\
    &\leq \lim_{T\to\infty}\frac{1}{T^2} \left[T\right]^2\left[m~2^{2N}\right]\\
    &\leq m~2^{2N}.
\end{align*}

The first inequality is due to $\sum_{j=1}^m (t_j^f - t_j^i)k_j>1$, the second due to the Cauchy-Schwarz inequality, the third due to $t_j^f - t_j^i>0$ and $k_j\in\{1, 2, \dots, 2^N\}$, and the fourth because $\sum_{j=1}^m (t_j^f - t_j^i)\leq T$. Therefore, $\Omega\leq m~2^{2N}$, where $N$ is the number of nodes in the network and $m$ the number of attractors reached during the dynamics. Observe that, as $m\to\infty$, $\Omega\to\infty$. This is again coherent with our conception of open-endedness, as generating an unbounded number of potential phenotypes is translated as being open-ended.~\cite{stock2024open} In the Discussion section, we have situated our metric with respect to previous definitions and measurement programs in the literature on OEE, as well as provided an extension of $\Omega$ to encompass continuous state spaces.

\section{Data Availability}
The minimal dataset required to interpret, replicate and build upon the findings reported in this article, including the data used to generate the curves, is openly available in the following GitHub repository: \url{https://github.com/amahury/OEE-metric/tree/main/data/npy} (accessed on 8 April 2026).

\section{Code Availability}
The code used to perform the simulations in this study is openly available in the following GitHub repository: \url{https://github.com/amahury/OEE-metric} (accessed on 8 April 2026).

\section{Acknowledgments}
The authors acknowledge the support of the SSIE School (NYS 910247-22) to cover article processing charges for the present manuscript.

\section{Author Contributions}
A.J.L.‑D. conceptualized the project, carried out formal analysis, performed simulations, and wrote the original draft. P. J. R. T. provided theory and code related to PBN. G. L. F. developed the extension of the metric to continuous/hybrid state spaces. C.G. acquired funding, provided computational resources, and supervised. All authors validated, reviewed, edited, and wrote the final manuscript.

\section{Competing Interests}
The authors declare no competing interests.

\footnotesize
\bibliographystyle{naturemag}
\bibliography{example} 

\end{document}